\documentstyle[12pt]{article}
\begin{document}
\begin{flushleft}{\Large \bf Bethe Ansatz for the Spin-1 XXX Chain with
                    Two Impurities}
\end{flushleft}
\vspace{5mm}
\begin{flushleft}
 \hspace{0.5cm}HOU Boyu, SHI Kangjie, YUE Ruihong and ZHAO Shaoyou\\
\vspace{4mm}
  \hspace{0.5cm}{\small Institute of modern physics, Northwest university,
P.O.Box 105}\\
\hspace{0.5cm}{\small Xi'an 710069, P.R. China}\\
\vspace{4mm}
 \hspace{0.5cm}{\bf Abstract} {\sl By using algebraic Bethe ansatz
method, we give the Hamitonian of the\\
  \hspace{0.5cm}spin-1 XXX chain associated with  $sl_2$  with two
boundary impurities}.\\
\vspace{10mm}
\hspace{0.5cm}{\bf Pacs numbers:} 75.20H, 12.40E, 75.10D\\
 \hspace{0.5cm}{\bf Key words:} Algebraic Bethe ansatz, Impurity, Spin-1 XXX
Heisenberg chain\\
 \end{flushleft}
\section{Introduction}

   It is well known that the spin dynamics of the Kondo problem is equivalent
to the dynamics of the spin chain with magnetic impurities$^{\cite{NK}}$. Much
progress has been made by using 	 renomalization-group
technique, conformal field theory and exact diagonalization method(Ref.[2-9]).

     Magnetic impurities play an important role in the one-dimensional
 strongly correlated electron systems. However, the impurity  usually destroys the
integrability  when it is embedded into the "pure"quantum chain. So it is a
challenging problem to deal with impurity effects in integrable models 
without losing the integrability.The quantum inverse scattering method and 
the Bethe ansatz technique have been powerful tools to study  1-dimensional 
integrable impurity problems, such as supersymmetric
 $U$ model$^{\cite{Zhou}}$ and $t-j$ model with impurities (Ref.[6-9]).

   Andrei and Johannesson  were the first group who  considered the integrable
 impurity problem of the Heisenberg chain with periodic boundary
condition$^{\cite{An}}$ .
 Then Lee and Schlottmann generalized their result to arbitrary
 spin impurity$^{\cite{Lee,SMan}}$. But They have to present some unphysical 
terms in Hamiltonian to maintain the integrability, though these terms may be
irrelevant$^{\cite{W}}$. Noticing
the fact that the open boundary theory is very useful to formulate both the
 thermodynamics and the transport properties of the quantum chain with
 impurities, Gaudin considered the nonlinear Schr${\ddot o}$dinger model and
the spin-1/2 Heisenberg chain with simple open boundary$^{\cite{Gan}}$,
 then Schulz and Alcaraz$^{\cite{Sch,AB}}${\sl et al.} generalized it to Hubbard
and other models. 
In recent researches,   the spin-$1/2$ XXX chain with arbitrary spin 
impurities$^{\cite{W}}$ and spin-$1/2$ XXZ chain with  spin-1/2
impurities$^{\cite{Hupu}}$ have been discussed.	

     In this paper we consider the impurity effects of the 19-vertex model which is
associated with spin-1 representation of $sl_2$. This model is the simplest
 extension of the six-vertex model. Yue have obtained the precise elliptic 
solution of the Yang-Baxter equation of the spin-1 Heisenberg chain$^{\cite{Y}}$
by using the fusion procedure. In this paper we only use the rational limit of
the solution.

   The outline of this paper is as follows. In Section 2 we introduce the
19-vertex model briefly and discuss the integrability of 19-vertex with
impurities. Then  we construct the transfer matrix and its eigenvalues 
in the frame of  algebraic Bethe ansatz in section 3. A brief discussion
about the results is given  in the last section.

    \section{The Spin-1 XXX Heisenberg Chain with Impurities}

     The R-matrix of ordinary spin-1 XXX Heisenberg chain can be derived by
  using the fusion method$^{\cite{Y,Zam}}$. It takes the form
 
\begin{equation}
R(u)=\rho(u)\left ( 
\begin{array}{ccccccccc}
a_1 & 0 & 0 & 0 & 0 & 0 & 0 & 0 & 0\\
0 & a_2 & 0 & a_3 & 0 & 0 & 0 & 0 & 0\\
0 & 0 & a_4 & 0 &a_5 & 0 & a_6 & 0 & 0\\
0 & a_3 & 0 & a_2 & 0 & 0 & 0 & 0 & 0\\
0 & 0 & a_5 & 0 & a_7 & 0 & a_5 & 0 & 0\\
0 & 0 & 0 & 0 & 0 & a_2 & 0 & a_3 & 0\\
0 & 0 & a_6 & 0 & a_5 & 0 & a_4 & 0 & 0\\
0 & 0 & 0 & 0 & 0 & a_3 & 0 & a_2 & 0\\
0 & 0 & 0 & 0 & 0 & 0 & 0 & 0 & a_1
\end{array} \right ),
\end{equation}
  where
$$ 
 \begin{array}{llll}
  \rho(u)=1/(u+2), &  a_1=u+2,   &   a_2=u,    &  a_3=2, \\
   a_4=u(u-1)/(u+1),&   a_5=2u/(u+1), & a_6=2/(u+1), & a_7=u+a_6.
 \end{array}
$$
  It satisfies the Yang-Baxer equation
\begin{equation}
      R_{12}(u-v)\stackrel{1}{T}(u)\stackrel{2}{T}(v)
    =\stackrel{2}{T}(v)\stackrel{1}{T}(u)R_{12}(u-v),  
\end{equation}
 where $R_{12}$ acts on the auxiliary space $V\otimes V$. The $T$
 acts on quantum spaces as $v_1\otimes v_2\otimes\cdots
\otimes v_N$, with the notation $\stackrel{1}{A}=A\otimes 1$,
$\stackrel{2}{A}=1\otimes A$, and $T(u)$ is defined
by $$T(u)=\prod _{j=1}^{N}R_{ij}(u).$$

In order to construct the integrable 19-vertex model with  open boundary condition, 
we introduce the reflection  equation
 \begin{equation}
R_{12}(u-v)\stackrel{1}{K}(u)R_{21}(u+v)\stackrel{2}{K}(v)
         =\stackrel{2}{K}(v)R_{12}(u+v)\stackrel{1}{K}(u)R_{21}(u-v),
\end{equation}
where $K$ is the reflecting matrix which determines
the boundary terms in the Hamiltonian. One can prove that the double-row
monodromy
matrix defined by $U(u)=T(u)K(u)T^{-1}(-u)$  satisfies
the above reflection equation.
\begin{equation}
R_{12}(u-v)\stackrel{1}{U}(u)R_{21}(u+v)\stackrel{2}{U}(v)
          =\stackrel{2}{U}(v)R_{12}(u+v)\stackrel{1}{U}(u)R_{21}(u-v).
 \end{equation}

The dual $K$-matrix $K_+(u)$ can be defined by the automorphism$^{\cite{Sk}}$
 \begin{equation}
    \phi:  K(u)\rightarrow K_+(u)=K^t(-u-\eta).
 \end{equation}
It satisfies the dual refection equation
 \begin{eqnarray}
& &R_{12}(-u+v)\stackrel{1}{K}_+(u)R_{12}(-u-v-2\eta)\stackrel{2}{K}_+(v)\nonumber\\
&=&\;\;\stackrel{2}{K}_+(v)R_{12}(-u-v-2\eta)\stackrel{1}{K}_+(u)R_{12}(-u+v).
\end{eqnarray}
Then the transfer matrix can be defined by
\begin{equation}
t(u)=tr K_+(u)U(u),.
\end{equation}
One can check  that the transfer matrix satisfies the
commutative relation
\begin{equation}[t(u),t(v)]=0.\end{equation}

 Now we couple the spin-1 Heisenbser chain with two spin-1 impurities
which is located at both ends of the system.  The $L$-operator satisfying
Yang-Baxter relation takes the form  
\begin{equation}
 L_n(u)=\frac{  2(\vec{S}\cdot\vec{S}_n)^2
                 +2(u+1)\vec{S}\cdot\vec{S}_n
               +u^2+u-2}{(u+1)(u+2)}, \\
     \end{equation}
with  $n=1,2,\cdots,N,a,b$ and  $\vec{S}$ defined by 
 $$
     \begin{array}{lll}
    S^1=\frac{1}{\sqrt{2}}\left ( \begin{array}{ccc}
     0 & 1 & 0 \\
     1 & 0 & 1 \\
     0 & 1 & 0     \end{array} \right ), &
    S^2=\frac{1}{\sqrt{2}}\left (\begin{array}{ccc}
     0 & -i & 0 \\
     i & 0 & -i \\
     0 & i & 0      \end{array} \right ),&
    S^3=\left ( \begin{array}{ccc}
    1 & 0 & 0\\
    0 & 0 & 0\\
    0 & 0 & -1    \end{array} \right ).
   \end{array}
  $$
 The spin-1 operators $\vec{S}_n$ act on the $n$-th quantum space.  Then we define
 \begin{eqnarray}
  T(u)&=&L_{b}(u+c_b)L_{N}(u)L_{N-1}(u)\cdots
L_{2}(u)L_{1}(u),\nonumber\\
  \tilde{T}(u)&=&T^{-1}(-u)=L_{ 1}(u)L_{ 2}(u)\cdots L_{ N-1}(u)L_{
N}(u)L_{ b}(u-c_b),
\end{eqnarray}
where $c_b$ is arbitary constant.
 
  Then we put
\begin{eqnarray}
  &&K_+(u)=1,\nonumber\\
  &&K(u)=L_{a}(u+c_a)L_{
a}(u-c_a).\;\;\;\;(c_a\;\;is\;\;also\;\; 
a\;\;constant) 
  \end{eqnarray}
  Obviously $K(u)$ and $K_+(u)$ satisfy the reflection equation (3) and the dual
reflection eqution (6) respectively . One can check that they also
satisfy the commutative relation(8).

 \par From equation (8), we can obtain infinit number of conserved quantities by
expanding $t(u)$
in terms of $u$, which ensures the integrability of this model. The
Hamiltonian of this model can be written as 
 \begin{equation}
 H=\frac{d}{du}tr_{\tau}K_+U(u)\mid_{u=0}.
 \end{equation}
 After some calculation, we obtain the precise result of the Hamiltonian
up to a constant 
\begin{eqnarray}
H&=&\sum_{j=1}^{N-1}\left [-(\vec{s}_j\cdot\vec{s}_{j+1})^2
           +\vec{s}_j\cdot\vec{s}_{j+1}\right ]\nonumber\\
  &&         +J_a \left
        [(\vec{s}_1\cdot\vec{s}_{a})^2+(c_a^2-1)\vec{s}_1\cdot\vec{s}_{a}\right ]
           +J_b \left
        [(\vec{s}_N\cdot\vec{s}_{b})^2+(c_b^2-1)\vec{s}_N\cdot\vec{s}_{b}\right ],
\end{eqnarray}
where
\begin{eqnarray*}
J_i=-\frac{4}{(c_i^2-4)(c_i^2-1)},\;\;\;\;(i=a,b).
\end{eqnarray*}

\section{Algebraic Bethe Ansatz for the Impurities Model}

As usual, we rewrite the monodyomy matrix as
 \begin{equation}
  U(u)= \left ( \begin{array}{ccc}
           {\cal A}(u)   &  {\cal B} _1(u)  &  {\cal B} _2(u) \\
           {\cal C}_1(u)  &  {\cal D} _1(u)  &  {\cal B} _3(u) \\
           {\cal C}_2(u)  &  {\cal C} _3(u)  &  {\cal D} _2(u)   
\end{array} \right ),
 \end{equation}
and define the  vacuum state by
\begin{equation}
   \vert 0\rangle=\prod_{j=a,1}^{N,b}\otimes \left ( \begin{array}{l}
              1 \\ 0 \\ 0 \end{array} \right ).
  \end{equation}
 Applying $U(u)$ on the vacuum state, one can directly check
\begin{equation}  
\begin{array}{ll}
  {\cal B}_i(u)\vert 0 \rangle \ne 0, &  
  {\cal C}_i(u)\vert 0 \rangle =0. \;\;\;(i=1,2,3)
\end{array}
\end{equation}
In order to simplify our calculation, we introduce the transformation
\begin{eqnarray}
{\hat{\cal D}}_1(u)&=&{\cal D}_1(u)-\frac{1}{u+1}{\cal A}(u),\nonumber\\
{\hat{\cal D}}_2(u)&=&{\cal D}_2(u)-\frac{1}{u}{\hat{\cal D}}_1(u)
                -\frac{1}{(u+1)(2u+1)}{\cal A}(u).
\end{eqnarray}
Under the new notation, the transfer matrix can be written as
  \begin{eqnarray}
  t(u) &=&{\cal A}(u)+{\cal D}_1(u)+{\cal D}_2(u)\nonumber\\
       &=&w^+_1{\cal A}(u)+w_2^+{\hat{\cal D}}_1(u)+w_3^+{\hat{\cal D}}_2(u)\nonumber\\
       &=&\frac{2u+3}{2u+1}{\cal A}(u)+\frac{u+1}{u}{\hat{\cal D}}_1(u)
            +{\hat{\cal D}}_2(u).
  \end{eqnarray}
   where we have introduced $w_i^+ (i=1,\;\;2,\;\;3)$ for convenience.
Then, we can find the eigenvalues of 
     ${\cal A}(u),{\hat{\cal D}}_1(u),{\hat{\cal D}}_2(u)$ as
 \begin{equation}
 \begin{array}{rr}
 {\cal A}(u)\vert 0\rangle=w_1\vert 0\rangle, &
 {\hat{\cal D}}_1(u)\vert 0\rangle=w_2\vert 0\rangle,\\
 {\hat{\cal D}}_2(u)\vert 0\rangle=w_2\vert 0\rangle,
 \end{array}
 \end{equation}
 where \begin{eqnarray*}
 w_1&=&1,\\
w_2&=&\frac{u}{u+1}\cdot\frac{u^{2N}}{(u+2)^{2N}}\cdot\frac{(u^2-c_a^2)(u^2-c_b^2)}
          {((u+2)^2-c_a^2)((u+2)^2-c_b^2 )}, \\
 w_3&=&\frac{2u-1}{2u+1}\cdot\frac{[u(u-1)]^{2N}}{[(u+1)(u+2)]^{2N}}\cdot\\
     &&
\frac{(u^2-c_a^2)(u^2-c_b^2)[(u-1)^2-c_a^2][(u-1)^2-c_b^2]}
               {[(u+1)^2-c_a^2][(u+1)^2-c_b^2][(u+2)^2-c_a^2]
                          [(u+2)^2-c_b^2]}. 
      \end{eqnarray*}

\par From the reflection equation (4), we have
 \begin{eqnarray}
 a_1^-a_2^+{\cal B}_1(u){\cal A}(v)
         &=&a_1^+a_2^-{\cal A}(u){\cal B}_1(v)\nonumber\\
         & &\mbox{}+a_3^+a_2^-{\cal B}_1(u){\cal D}_1(v)+a_2^
                   +a_3^-{\cal B}_1(u){\cal A}(v),
 \end{eqnarray}
with the nonations $a_i^{\pm}=a_i(u{\pm}v),(i=1,2,\cdots,7).$ Replacing
${\cal D}_1$ with ${\hat{\cal D}}_1$ ,we find the commutative relation of
the ${\cal A}(u)$ and ${\cal B}_1(v)$ on vacuum state,
\begin{eqnarray}
  {\cal A}(u){\cal B}_1(v) \vert 0 \rangle
    &=&\frac{(u+v)(u-v-2)}{(u-v)(u+v+2)}{\cal B}_1(v){\cal A}(u) \vert 0 
                                                           \rangle\nonumber\\
    & &\mbox{}-\frac{2}{u+v+2}{\cal B}_1(u){\hat{\cal D}}_1(v) \vert 0  
                                                            \rangle\nonumber\\
    & &\mbox{}+\frac{2v}{(v+1)(u-v)}{\cal B}_1(u){\cal A}(v) \vert 0 \rangle.
\end{eqnarray}
Similarly, we also find the following two equations
\begin{equation}
\begin{array}{l}
{\cal A}(u){\cal B}_1(v)a_3^+a_3^-+{\cal D}_1(u){\cal B}_1(v)a_2^+a_2^-
    +{\cal B}_1(u){\cal D}_1(v)a_7^+a_3^- \\
\;\;\;\;\mbox{}    +{\cal B}_3(u){\cal D}_1(v)a_5^+a_2^- \\
={\cal B}_1(v){\cal A}(u)a_5^+a_5^-+{\cal A}(v){\cal B}_1(u)a_3^+a_7^-
     +{\cal B}_1(v){\cal D}_1(u)a_7^+a_7^-\\
\;\;\;\;\mbox{}+{\cal A}(v){\cal B}_3(u)a_2^+a_5^-+{\cal B}_1(v){\cal D}_2(u)a_5^+a_5^- ,
\end{array}\end{equation}
\begin{equation}\begin{array}{l}
{\cal B}_1(u){\cal A}(v)a_5^+a_3^-+{\cal B}_3(u){\cal A}(u)a_4^+a_2^-\\
={\cal B}_1(v){\cal A}(u)a_5^+a_6^-+{\cal A}(v){\cal B}_1(u)a_3^+a_5^-
          +{\cal B}_1(v){\cal D}_1(u)a_7^+a_5^-\\
\;\;\;\;\mbox{}+{\cal A}(v){\cal B}_3(u)a_2^+a_4^-+{\cal B}_1(v){\cal D}_2(u)a_5^+a_4^-.
\end{array}\end{equation}
Solving the above equations and substituting the commutative relation of
${\cal A}$
 and ${\cal B}_1$ into it, we obtain the commutative relation of ${\hat{\cal
D}}_1$ and ${\cal B}_1$ on the vacuum state,
\begin{eqnarray}
{\hat{\cal D}}_1(u){\cal B}_1(v) \vert 0 \rangle
       & =&\frac{(u+v)(u-v+1)(u-v-2)(u+v+3)}{(u-v)(u+v+1)(u-v-1)(u+v+2)}{\cal
                       B}_1(v){\hat{\cal D}}_1(u) \vert 0 \rangle\nonumber\\
          &&\mbox{}-\frac{2\left [ u(u+v)+(u+2v)\right ]}{(u+1)(u-v)(u+v+1)}\;{\cal
                       B}_1(u){\hat{\cal D}}_1(v) \vert 0 \rangle\nonumber\\
          &&\mbox{}+\frac{2v\left [u(u-v)-(u+2v+4)\right ]}{(u+1)(v+1)(u-v-1)(u+v+2)}{\cal
                       B}_1(u){\cal A}(v) \vert 0 \rangle\\
          &&\mbox{}-\frac{2}{u+v+1}{\cal B}_3(u){\hat{\cal D}}_1(v) \vert 0
                                                             \rangle
          +\frac{2v}{(v+1)(u-v-1)}{\cal B}_3(u){\cal A}(v) \vert 0 \rangle.
                                                           \nonumber
 \end{eqnarray}
 Using the similar procedure we can find the third commutative relation
\begin{eqnarray}
 {\hat{\cal D}}_2(u){\cal B}_1(v) \vert 0 \rangle
&=&\frac{(u-v+1)(u+v+3)}{(u-v-1)(u+v+1)}{\cal B}_1(v){\hat{\cal D}}_2(u)\vert 0
\rangle\nonumber\\
& &\mbox{}+\frac{4(u+1)}{u(2u+1)(u-v-1)}{\cal B}_1(u){\hat{\cal D}}_1(v) \vert 0
\rangle\nonumber\\
& &\mbox{}-\frac{4v(u+1)}{u(2u+1)(v+1)(u+v+1)}{\cal B}_1(u){\cal A}(v) \vert 0 \rangle \\
& &\mbox{}-\frac{2(u+1)}{u(u-v-1)}{\cal B}_3(u){\hat{\cal D}}_1(v) \vert 0 \rangle
       +\frac{2v(u+1)}{u(v+1)(u+v+1)}{\cal B}_3(u){\cal A}(v)\vert 0
\rangle.\nonumber
 \end{eqnarray}

Applying the transfer matrix $t(u)$ on the one-particle state $\vert v\rangle={\cal
B}_1(v)\vert 0\rangle$, we have
\begin{eqnarray}
t(u)\vert v\rangle
     &=&\left\{w_1w_1^+\frac{(u+v)(u-v-2)}{(u-v)(u+v+2)}\right. \nonumber\\
     & &\mbox+w_2w_2^+\frac{(u+v)(u-v+1)(u-v-2)(u+v+3)}
                           {(u-v)(u+v+1)(u-v-1)(u+v+2)}\\
     & &\left.\mbox+w_3w_3^+\frac{(u-v+1)(u+v+3)}{(u-v-1)(u+v+1)}\right\}
                                                   \vert v\rangle \nonumber
\end{eqnarray}
under the following constrain
\begin{equation}
       \frac{(v+1)^{2N}[(v+1)^2-c_a^2][(v+1)^2-c_b^2]}
            {(v-1)^{2N}[(v-1)^2-c_a^2][(v-1)^2-c_b^2]}=-1.
\end{equation}

    From the reflection equation (4), we can construct the two-particle
excited state as 
   \begin{eqnarray}
   \vert v_1,v_2\rangle&=& \{{\cal B}_1(v_1){\cal B}_1(v_2)
          +\frac{2}{v_1+v2+1}{\cal B}_2(v_1){\cal D}_1(v_2)\nonumber\\
             &&\;\;-\;\frac{2(v_1+v_2-1)}{(v_1-v_2-1)(v_1+v_2+1)}{\cal
B}_2(v_1){\cal A}(v_2)\}\vert 0 \rangle,
  \end{eqnarray}
which is symmetric in $v_1,v_2$ up to a whole factor.
Applying the transfer matrix on the above state, one can find a lot of
``unwanted terms". To ensure the above state to be the eigenstate of the transfer
matrix, the ``unwanted terms" mush vanish. However, in present case, the
calculation become much more complicated that we expected. We have to apply the
functional Bethe ansatz method which was first proposed for the Ising
model$^{\cite{Bax}}$. The key idea is that the eignvalue is an entire function.
Generally, the eigenvalue has superficial poles. The analytic property of 
the  eigenvalue  requires its residues at these poles to be zero,
which give the Bethe ansatz equations. The fact that the results obtained here 
coincide with those  obtained by
other methods suggests that this Bethe ansatz method is sound.

Tarasov $\sl{et\; al}$ argued that the functional BA method valided in  two-particle
case can be generalized to m-particle excitated states (Ref.[18-22]). Applying the
transfer matrix to the m-particle excitated states $|v_1,\cdots,v_m\rangle$, we
obtain the eignvalue

\begin{eqnarray}
 t^m(u\vert v_1,v_2,\cdots,v_m)
&=&w_1w_1^+\prod_{i=1}^{m}\frac{(u+v_i)(u-v_i-2)}{(u-v_i)(u+v_i+2)} \nonumber\\
& &\mbox+w_2w_2^+\prod_{i=1}^{m}\frac{(u+v_i)(u-v_i+1)(u-v_i-2)(u+v_i+3)}{(u-v_i)
       (u+v_i+1)(u-v_i-1)(u+v_i+2)}\nonumber\\
& &\mbox+w_3w_3^+\prod_{i=1}^{m}\frac{(u-v_i+1)(u+v_i+3)}{(u-v_i-1)(u+v_i+1)},
\end{eqnarray}
where $v_1,\cdots,v_m$ should satisfy the following restrictions
\begin{equation}
\frac{(2v_j+1)(v_j+1)^{2N}[(v_j+1)^2-c_a^2][(v_j+1)^2-c_b^2]}
     {(2v_j-1)(v_j-1)^{2N}[(v_j-1)^2-c_a^2][(v_j-1)^2-c_b^2]}
        =\prod_{i=1}^{m}\frac{(v_j-v_i+1)(v_j+v_i+1)}{(v_j+v_i-1)(v_j-v_i-1)}.
\end{equation}
\nopagebreak[4]
\section{Discussion}

 With the algebraic Bethe ansatz method, we have obtained the exact diagonalization
 Hamiltonian of the spin-1 XXX Heisenberg chain with impurities. This result can be
 used to discuss the other properties of the system when impurities coulpe to
 the chain, such as the thermodynamics, the transport property {\sl et
 al}. The Hamiltonian of the other high spin Heisenberg chain can also be
 obtained by using the similar method. It is also a good course to
 study the arbitrary spin impurities  coupled to  the  Heisenberg
 chain. And it seems to bring us more information about impurity effects in strongly
 correlated electron systems.  
 
 \end{document}